%% file: glueFT_final.tex
\documentclass[aps,prl,twocolumn,showpacs,superscriptaddress,groupedaddress]{revtex4-1}  
\usepackage{graphicx}  
\usepackage{dcolumn}   
\usepackage{bm}        
\usepackage{amssymb}   

\newcommand{\eq}{\begin{equation}} 
\newcommand{\en}{\end{equation}}

\begin{document}



\title{Finite temperature behaviour of glueballs in Lattice Gauge Theories}
\input author_list.tex       
\date{\today}

\begin{abstract}
We propose a new method to compute glueball masses in finite temperature Lattice Gauge Theories which
at low temperature is fully compatible with the known zero temperature results and as the temperature increases
leads to a glueball spectrum which vanishes at the
deconfinement transition. We show that this definition is consistent with the Isgur-Paton model and with the expected contribution of the glueball spectrum to various
thermodynamic quantities at finite temperature. We test our proposal with a set of high precision numerical simulations in the 
3d gauge Ising model and find a good agreement with our predictions.
\end{abstract}

\maketitle

While the physics of glueballs in pure LGTs  at zero temperature 
is by now rather well understood \cite{Lucini:2004my,Lucini:2010nv,Gregory:2012hu} a similar level of
understanding for the finite temperature behaviour of the glueball spectrum is still lacking. 
The standard method used to compute finite $T$ glueball masses \cite{Ishii:2002ww,Meng:2009hh}  is to measure the correlator
of the glueball operator (for instance a simple plaquette, if one is interested in the $0^{++}$ glueball).
along the compactified time direction of length $ \frac{1}{T} $.
The glueball spectrum obtained in this way turns out to be
almost constant as the temperature increases and
seems not to be affected by the deconfinement transition: 
glueball masses were measured even deeply in the deconfined phase showing values similar to the zero
temperature ones (or slightly smaller, depending on the procedure adopted in the calculation)\cite{Ishii:2002ww,Meng:2009hh}.
However this picture is unsatisfactory for at least two reasons.

{\sl First}, one of the most successful phenomenological descriptions of glueballs is the well known Isgur-Paton model \cite{Isgur:1984bm}.
This model and its recent generalizations \cite{Johnson:2000qz}  is able not only to predict 
the general structure of the spectrum (like, for
instance, the
fact that the mass of the $2^{++}$ state is lower than the mass of the $1^{++}$) but also 
its fine details and shows a remarkable agreement with the lattice estimates. 
In this model glueballs are considered as "rings of glue", kept together by the same string tension
which appears in the interquark potential and should thus vanish at the deconfinement point when the string tension vanishes.
If we trust this picture, then it could be used also to predict the $T$ dependence of the glueball spectrum for low temperatures. 
In fact the Isgur Paton model predicts values of the zero temperature
glueball masses $m_i(T=0)$ as adimensional ratios $m_i(0)/\sqrt{\sigma(0)}$ (where $\sigma(0)$ is the zero temperature string tension).
In a finite temperature setting we expect the same ratios, but with $\sigma(0)$ substituted by the finite temperature string tension
$\sigma(T)$ i.e.:

\eq
m_i(T)=\frac{m_i(0)}{\sqrt{\sigma(0)}}\sqrt{\sigma(T)} 
\label{eq1}
\en

\noindent
which is a {\sl decreasing} function of $T$. This expectation is in complete disagreement with the almost constant
$T$ dependence proposed in \cite{Ishii:2002ww,Meng:2009hh}  for this range of temperatures.

{\sl Second}, recently, very precise estimates of various thermodynamic
quantities have been obtained both below and above $T_c$.  in pure lattice gauge theories in 
$d=3+1$ \cite{Meyer:2009tq,Panero:2009tv} and in $d=2+1$ \cite{Caselle:2011fy,Caselle:2011mn} dimensions. For $T<T_c$ the thermodynamics of these theories is very well described
in terms of a gas of glueballs which are the only 
degrees of freedom of the theory in this regime. For $T>T_c$ the thermodynamics is well described by a gas of free gluons (and accordingly
the thermodynamic observables scale as $N^2$). If glueballs were present also in the deconfined phase they would give an additional
contribution to the thermodynamic observables, leading to results fully incompatible with the lattice measurements.  

These observations suggest that with the current method to extract finite $T$ glueball masses \cite{Ishii:2002ww,Meng:2009hh} one is probably
measuring some other finite size scale of the model whose relation with the glueball spectrum is similar to
the relation which exists between the spacelike string tension $\sigma_s$ and the finite temperature string tension $\sigma(T)$.
Indeed also $\sigma_s$, which is extracted from spacelike Wilson loops, is almost constant for $T<T_c$, increases for $T>T_c$ and it is well known to be completely 
unrelated to the finite temperature string tension $\sigma(T)$ which is instead extracted from Polyakov loop correlators.

\begin{figure}
\vskip -2.9cm
\includegraphics[scale=0.3]{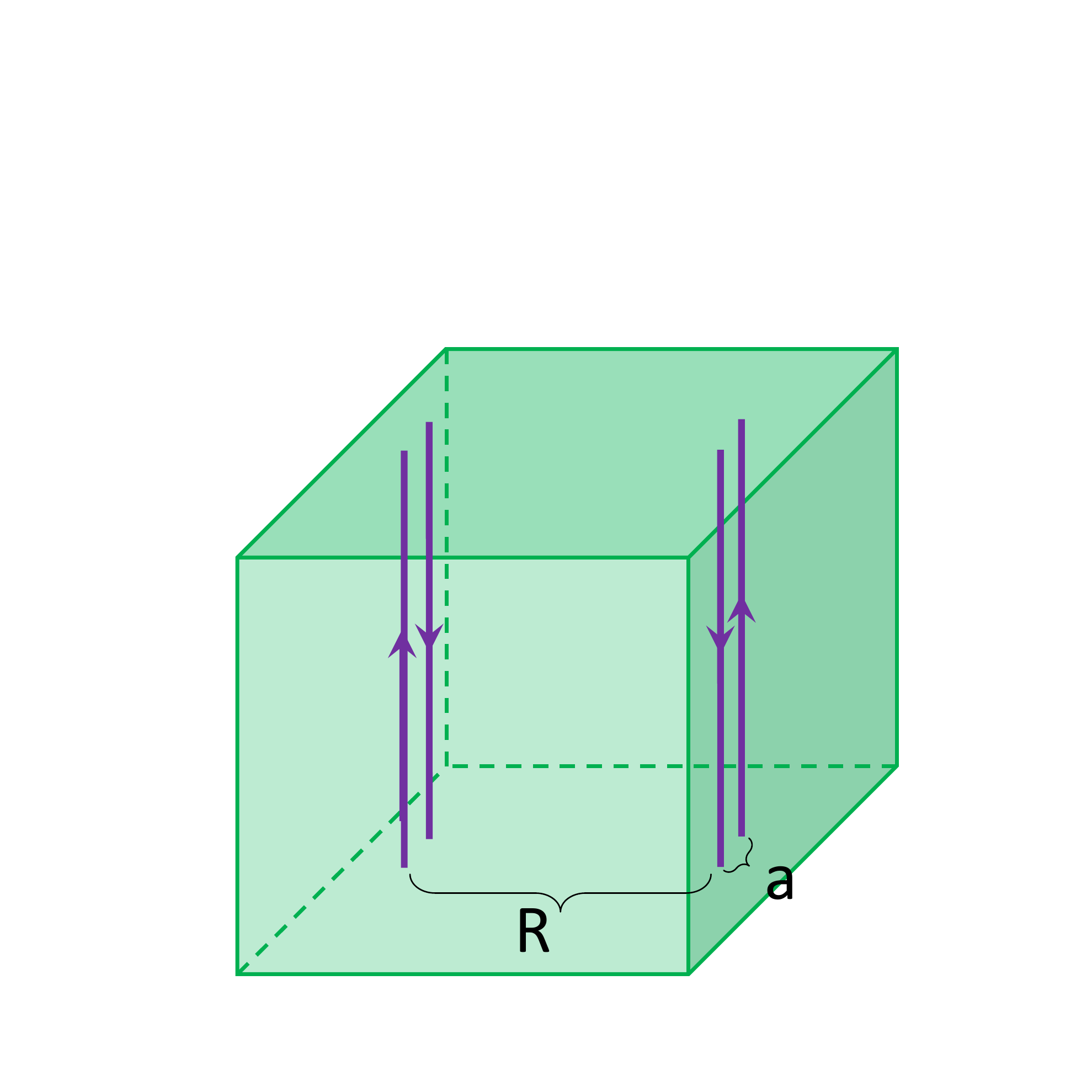}
\caption{\label{fig1} The glueball correlator discussed in the text.}
\end{figure}

In this letter we propose an alternative prescription to evaluate finite $T$ glueball masses,
compatible with the above observations. 
The most direct way to ensure the expected finite $T$ behaviour is to construct an observable 
with the correct quantum numbers so as to be coupled in the continuum limit to the glueball states, using only
Polyakov loops so as to have the correct dependence on the finite temperature string tension. 
The simplest proposal is to choose a pair  $P P^{\dagger}$ of nearby Polyakov loops
\begin{equation}
 M(x)=P(x)P^{\dagger}(x+a)
\end{equation}
where $a$ denotes the lattice spacing and $ P(x) $ is the Polyakov loop located at the space point $ x $. 
Then the glueball mass will be extracted looking at the large $R$ behaviour of the connected correlator of two $M(x)$ operators 
as depicted in fig.\ref{fig1}. 
\begin{equation}
G(R,T)\equiv  \langle M(0) M(R) \rangle - \langle M\rangle^2  \stackrel{\sim}{\tiny{}_{{ R \rightarrow \infty}}} e^{ -m_{0}(T) R }
\label{eq2}
\end{equation}
The space-like version (usually denoted as "torelon pair") was used in \cite{Gregory:2012hu,Lucini:2010nv} as part of the operator basis to obtain the $T=0$ 
glueball spectrum. In the proposed interpretation, this set up is new.

Let us also stress that this is not the only possible choice, for instance in non-abelian gauge theories an equivalent interesting possibility would be the Wilson loop obtained
joining the two Polyakov loops with two space like links at $t=0$ and $t=N_t$ (this choice is obviously equivalent to our proposal for abelian LGTs). 

\begin{figure}
\vskip -2.5cm
\includegraphics[scale=0.3]{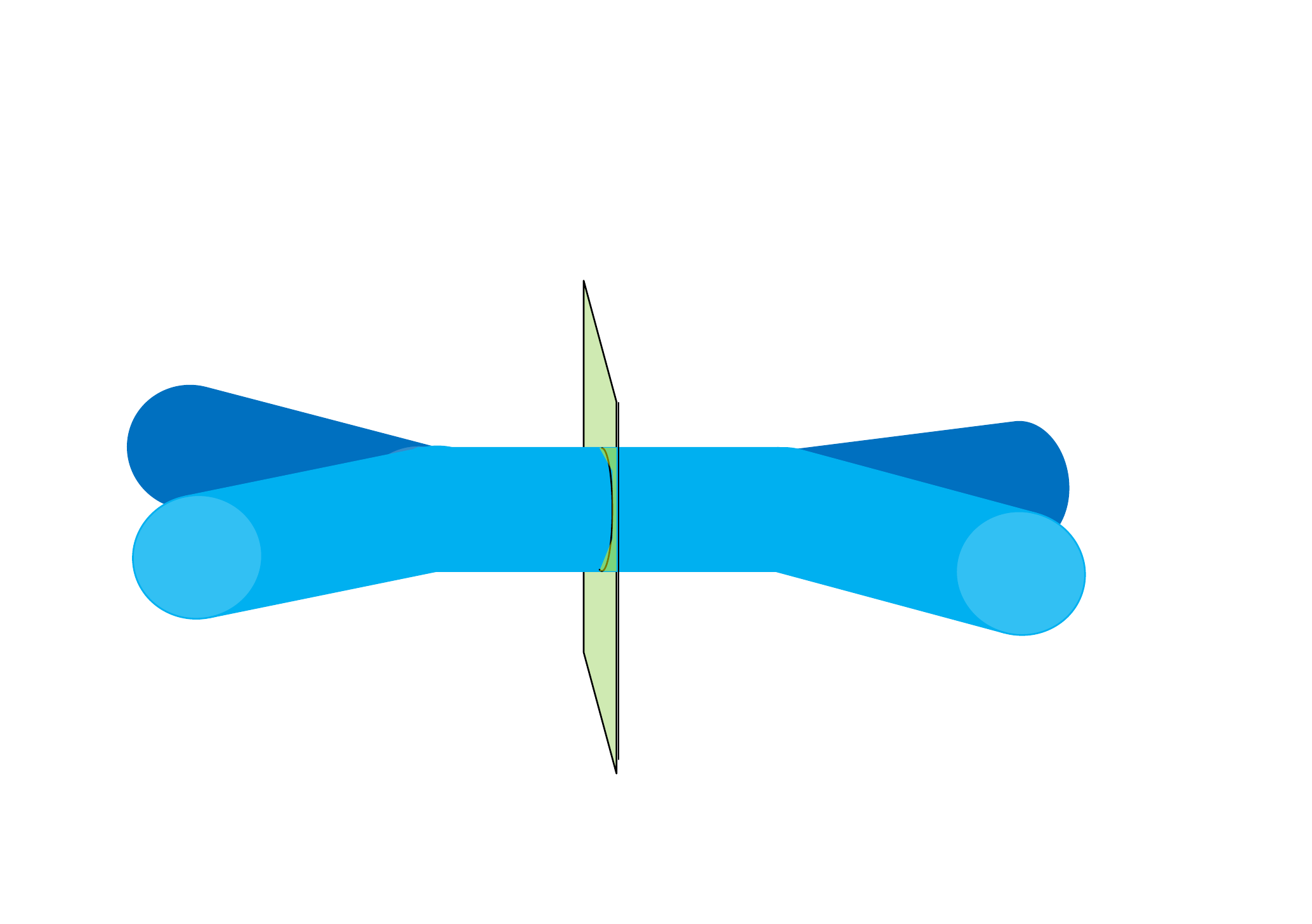}
\caption{\label{fig2} Effective string description of our proposal.}
\end{figure}

The nice feature of our proposal is that it has a natural interpretation in terms of the effective string model of pure gauge theories. It is the four point correlator
of four closed effective strings (see fig. \ref{fig2}). The external legs correspond to the four Polyakov loops 
(which are described as closed strings due to
the compactification of the time direction), while the
glueballs are the excitations of the closed string joining together the four legs.  
As mentioned in the introduction this proposal is strongly based on our intuition of the glueball dynamics 
coming from the Isgur-Paton model. It might be useful to make more explicit this connection. 
If we could perform a section in the four strings function as depicted in fig.2, the effective string description of the
flux distribution within the section would be given by the Isgur Paton model. Accordingly we expect that all the glueballs (independently
of their quantum numbers) would flow within the closed string, of which they would represent
different radial or rotational excitations. In the large $R$ limit only the lowest mass survives, but in principle, looking at the
subleading exponentials for lower values of $R$ one could recover also the remaining states of the spectrum. 
While the radius of the four external legs is fixed to be the inverse of the finite temperature,
the radius of the internal closed string coincides with the glueball radius $r_0$ which is one of the parameters of the Isgur-Paton model. 

Despite all these interesting features there is apparently a major problem with this proposal. In fact, due to dimensional reduction, one expects  
that any mass scale extracted from an observable of this type should scale in the vicinity of the deconfinement transition as  $m_s(T)\sim\sigma(T)/T$ 
which, as it is easy to see, is pretty different from the expected scaling behaviour of
eq.(\ref{eq1}). 

We shall see below in a concrete example how this problem can be addressed. Indeed, as we shall see, for any $T<T_c$, a (glueball) mass with the correct scaling
behaviour is always present in the spectrum of $G(R,T)$, but
as the deconfinement transition is approached this mass becomes subleading and the large $R$ behaviour is dominated by a different mass scale (whose physical meaning we
shall discuss below) with the "wrong" scaling behaviour $m_s(T)\sim\sigma(T)/T$. However, even in the vicinity of $T_c$, there is always a suitable range 
of values of $R$ in which the glueball mass, even if subleading, can be unambiguously observed.

\subsection{Test in the 3d gauge Ising model}

In order to test our proposal we computed the mass of the lightest glueball 
in the 3d gauge Ising model. This choice has two relevant advantages.

{\sl First}, very precise estimates exist for the zero temperature
spectrum~\cite{Agostini:1996xy} with which we can compare our results in the low $T$ limit. In particular we know that in the range of $\beta$ values that we study in this
paper 
$m_0=3.15(5)\sqrt{\sigma(0)}$~\cite{Agostini:1996xy}. 

{\sl Second}, using dimensional reduction~\cite{sy82} and the fact that the 3d gauge Ising model has a second
order deconfinement transition in the same universality class of the 2d Ising magnetization transition, we can predict
the behaviour of the correlator $G(R,T)$ in the vicinity of $T_c$ using results borrowed from the exact solution
of the 2d Ising model. In this limit our observable becomes equivalent to the energy-energy correlator in the high temperature phase of the 2d Ising 
spin model. From the exact solution of the model we know that the large $R$ behaviour of the function should be dominated by  a new mass scale $m_s$ 
which is exactly twice the fundamental mass of the
model, which, from dimensional reduction, is known to be $\sigma(T)/T$ (see for instance the discussion in sect.2.3 of \cite{Caselle:2006wr}).
Thus, as anticipated, we expect in this limit  
\begin{equation}
m_s(T)\sim 2\sigma(T)/T
\label{scaling2}
\end{equation}

We performed three sets of simulations at  different values of the gauge coupling (corresponding to $\frac{1}{T_c}=5.67a$, $8a$ and $12a$ respectively \cite{Caselle:1995wn})
in order to test scaling corrections.
For each value of $\beta$ we chose a value of the lattice size in the spatial direction $L_s$ large enough to make finite size effects negligible, and 
studied various values of compactified time direction $N_t\equiv 1/T$ in the range $T<T_c$. For each value of $T$ we
 evaluated the correlator $G(R,T)$ for several values of $R$. We also evaluated for each $T$ 
in a separate simulation the finite temperature string tension
$\sigma(T)$ (using the methods discussed in \cite{Caselle:2002ah}) so as to be able to construct the scaling functions eqs. (\ref{eq1}) and (\ref{scaling2}). 
A few details on the simulations are reported in tab. \ref{datisim}.

\begin{table}[ht]
\centering
\begin{tabular}{ |c|c|c|c|c| }
\hline
$\beta$ & $\frac{1}{T_c}$ &  $ L_s $ &  $N_t$   & $ R $  \\
\hline
0.743543  & 5.67 a & 90 & 7,8,9 & $6 \leq R \leq 20$ \\
0.751805  & 8  a & 90 & 9,10,11,12,13,14,20,56,64 & $8 \leq R \leq 22$ \\
0.756427  & 12 a & 120 & 20 & $12 \leq R \leq 33$ \\
\hline
\end{tabular}
\caption{For each of the three  $\beta$ values we report the corresponding critical temperature $T_c$ and the values of  $L_s$, $N_t$  and $R$  
that we studied. \label{datisim}}
\end{table}

\begin{figure}
\vskip -1.1cm
\includegraphics[scale=0.4]{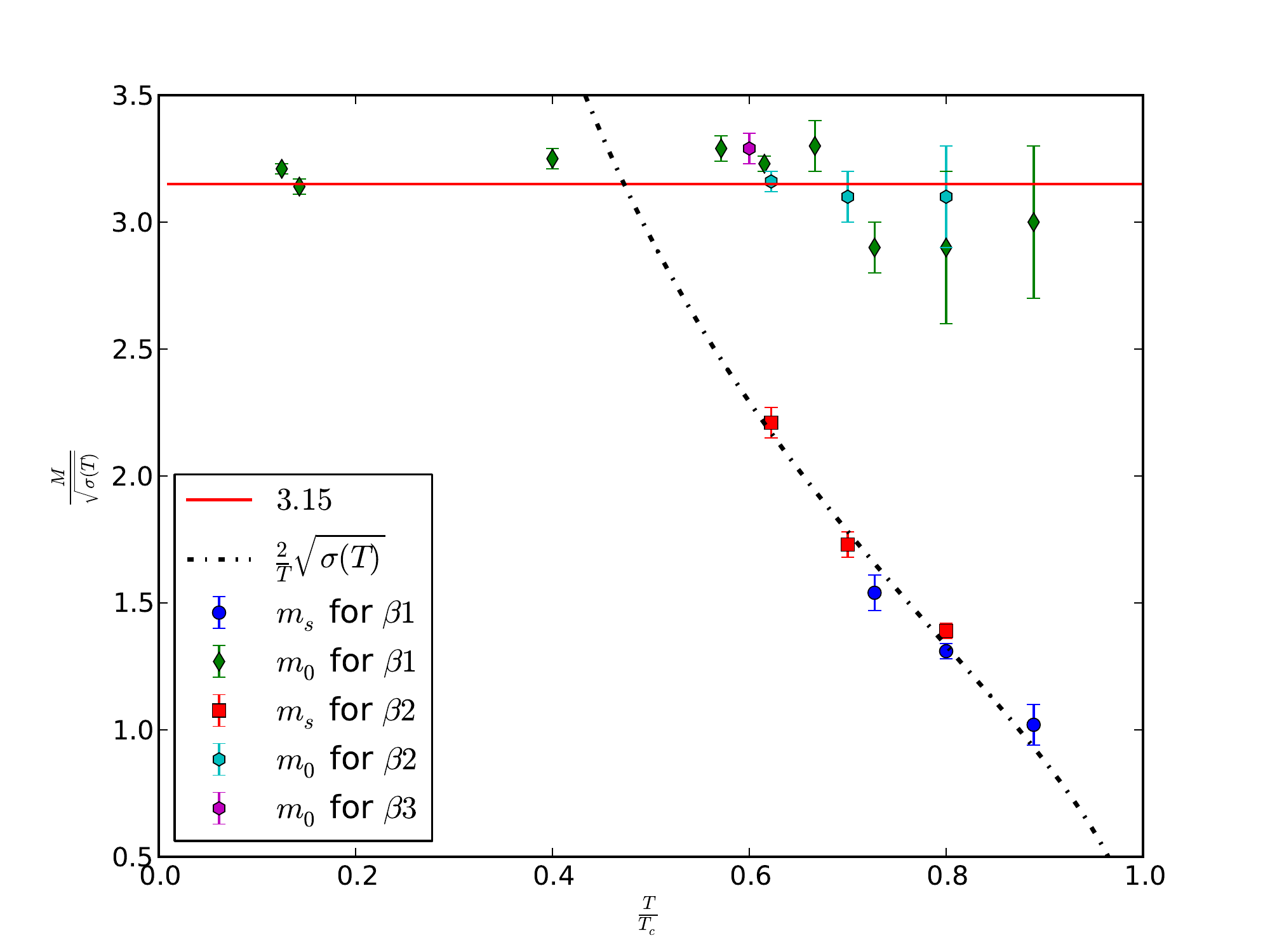}
\caption{\label{plot} $ \frac{m_0}{\sqrt{\sigma(T)}}$ and $\frac{m_{s}}{\sqrt{\sigma(T)}}$ plotted as a function of  $\frac{T}{T_c}$ for $\beta_1=0.743543$,
$\beta_2=0.751805$ and $\beta_3=0.756427$. The two curves 
correspond to the two expected scaling behaviours: $m_0(T)\sim 3.15 \sqrt{\sigma(T)}$ and $m_s(T)=2\sigma(T)/T$.}
\end{figure}

We found two different behaviours. For low values of $T$ (in our simulations the threshold was $T \lesssim 0.6 T_c$) 
the data were perfectly fitted by the following expression
\begin{equation}
G(R,T)=a_0(T) \frac{e^{-m_0(T) R}}{\sqrt{R}}
\label{eqfit1}
\end{equation}
with good $\chi^2$ values in the whole range of values of $R$ that we considered.
The data were so precise that we could also confirm the presence of the expected $1/\sqrt{R}$ prefactor. 
 The values of $m_0(T)$ extracted from the fits are reported in the last few lines (for each $\beta$) of tab.\ref{tabellamasse} and in fig.\ref{plot} and
turned out to follow exactly the expected behaviour eq.(\ref{eq1}), with a value of the glueball mass in good agreement with the $T=0$ value 
$m_0\sim 3.15\sqrt{\sigma(0)}$ obtained in~\cite{Agostini:1996xy}. 

For high values of $T$ (i.e. in our case $T \gtrsim 0.6 T_c$) it turned out to be impossible to fit the data using eq.(\ref{eqfit1}). Reasonable 
$\chi^2$ values could only be obtained discarding the low $R$ values of the
correlators and using a different fitting function:
\begin{equation}
G(R,T)=a_s(T) \frac{e^{-m_s(T) R}}{{R}^2}
\label{eqfit2}
\end{equation}
We use the notation $m_s$
 to stress the fact that this mass was obtained using a different fitting function. The prefactor $1/R^2$ is exactly what one would expect for the energy-energy correlator
 in the 2d Ising model and the mass $m_s$ extracted form this fit scales exactly as suggested by eq.(\ref{scaling2}). In full agreement with the expectation of
 dimensional reduction, not only the $T$ dependence but also the fact that $m_s$ is exactly $twice$ the fundamental mass of the model is perfectly reproduced by the data
 (see tab. \ref{tabellamasse} and fig.\ref{plot}).

This explains the large $R$ behaviour of $G(R,T)$, however in order to include also the small $R$ data in the fit it turned out to be mandatory to use a
two exponentials fitting function:
\begin{equation}
G(R,T)=a_0(T) \frac{e^{-m_0(T) R}}{\sqrt{R}}+a_s(T) \frac{e^{-m_s(T) R}}{{R}^2}
\label{eqfit3}
\end{equation}

It is well known that this type of fits is very delicate. In our case we used the following procedure: we kept the lowest mass $m_s$ fixed to the value obtained in the
large $R$ limit 
and fitted again the data keeping
as free parameters only $a_0(T)$, $a_s(T)$ and $m_0(T)$ we could in this way fit all the data with a reduced $\chi^2$ of order unity. 
It is important to stress that the identification of the subleading exponential was facilitated  by the very different behaviour of the two prefactors, by the wide range of
values of $R$ that we used in the fit and by the fact that these data were not cross-correlated since, due to the algorithm 
that we used (see \cite{Caselle:2002ah}), each value of $R$ was obtained in an independent simulation. These observations should be taken into account when trying to reproduce
our results in other LGTs.
The results of our fits are reported in tab. \ref{tabellamasse} and fig.\ref{plot}. The subleading mass $m_0$ in the 
$T>0.6 T_c$ region turned out to be the natural
 continuation of the glueball mass
that we had identified for $T<0.6 T_c$. As it is easy to see looking at fig.\ref{plot}, 
$m_0$ follows eq.(\ref{eq1}) up to the highest temperatures that we studied with a value
$m_0\sim 3.15$ in agreement with the $T=0$ estimate.

\begin{figure}
\vskip -2cm
\includegraphics[scale=0.3]{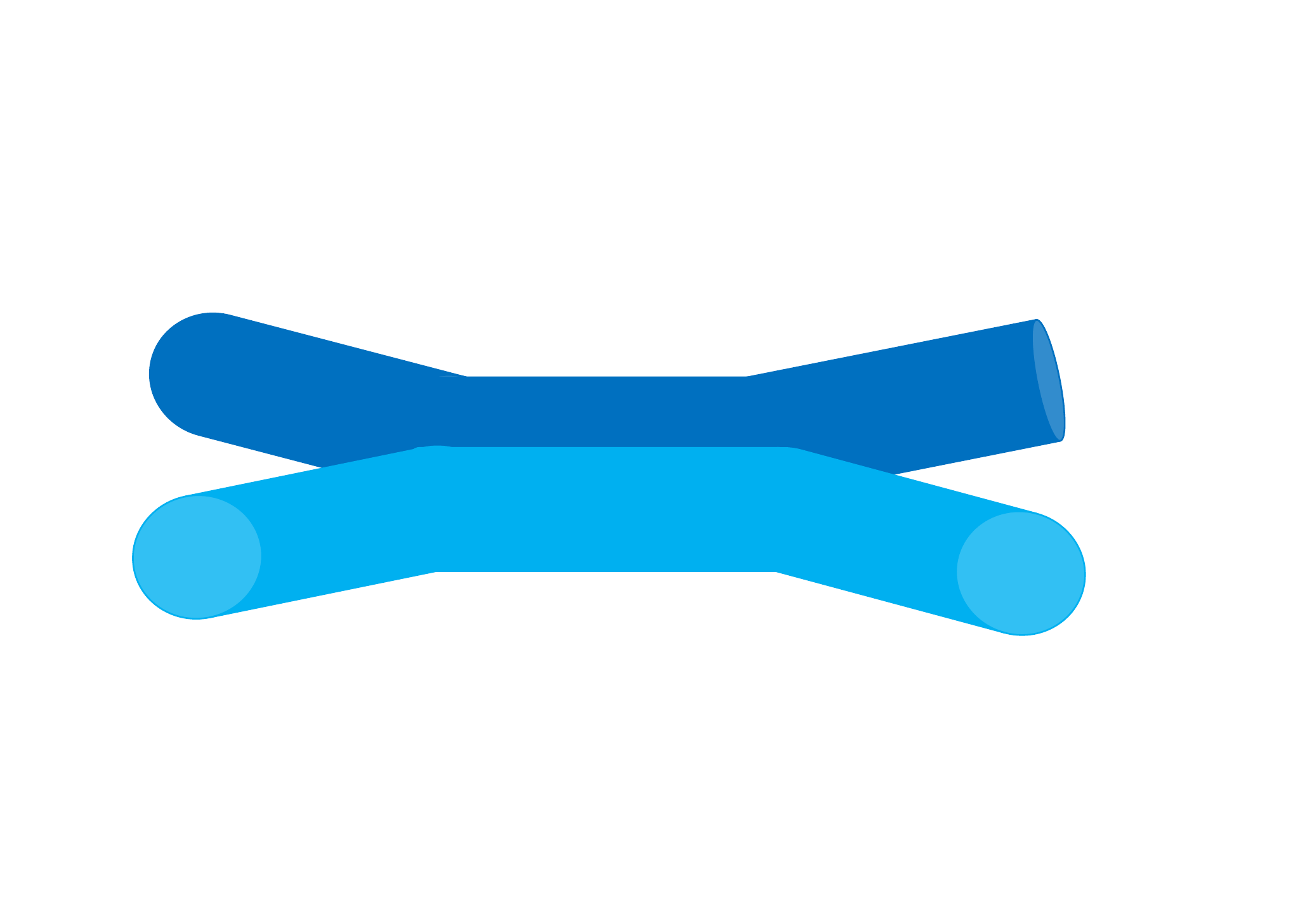}
\caption{\label{fig3} Minimal surface associated to the $m_s$ mass.}
\end{figure}

It would be important to understand the effective string interpretation of this new scale $m_s$.
We have no rigorous proof but it is likely that the crossover that we observe between $m_0$ and $m_s$
is due to a competition between the two  minimal surfaces bounded by the four Polyakov loops which are compatible with the topology of the lattice and of our observable.
 As the temperature increases it becomes less and less costly
for the flux tube to wind around the periodic boundary conditions leading to a minimal surface composed by two parallel 
flux tubes as depicted in fig.\ref{fig3} .  This crossover is controlled by the competition of two scales: the compactification radius 
$1/T$ and the glueball radius $r_0$. As $T$ increases also $r_0$ increases (since it is due to the flux tube width which is known to
increases with $T$) thus it will certainly exist a crossover value of $T$ above which 
$r_0>1/T$ which in the 3d gauge Ising model that we studied turns out to be around $T_c/2$.

\begin{table}[ht]
\centering
\begin{tabular}{ |cccccc| }
\hline
$\beta$ & $ \frac{T}{T_c} $ &  $ \sigma(T) $  & $ \frac{m_s(T) T}{\sigma(T)} $ & $ \frac{m_s(T)}{\sqrt{\sigma(T)}} $ & $ \frac{m_{0}(T)}{\sqrt{\sigma(T)}} $ \\
\hline
0.743543 & 0.8  & 0.00961 &  2.03(4)  & 1.39(3) & 3.1(2) \\
0.743543 & 0.7  & 0.01315 &  1.89(5)  & 1.73(5) & 3.1(1) \\
0.743543 & 0.62 & 0.01542 &  1.98(5)  & 2.21(6) & 3.14(8) \\
\hline
0.751805 & 0.89 & 0.00268 &  2.2(2)  & 1.02(8) & 3.0(3)  \\
0.751805 & 0.8  & 0.00444 &  1.97(5) & 1.31(3) & 2.9(3) \\
0.751805 & 0.73 & 0.00566 &  1.86(8) & 1.54(7) & 2.9(1) \\
0.751805 & 0.67 & 0.00654 &          &         & 3.3(1) \\
0.751805 & 0.62 & 0.00720 &          &         & 3.23(3) \\
0.751805 & 0.57 & 0.00771 &          &         & 3.29(5) \\
0.751805 & 0.4  & 0.00922 &          &         & 3.25(4) \\
0.751805 & 0.14 & 0.01037 &          &         & 3.14(3) \\
0.751805 & 0.125& 0.01040 &          &         & 3.21(2) \\
\hline
0.756427 & 0.6  & 0.00326 &          &         & 3.29(6)  \\
\hline
\end{tabular}
\caption{Values of $\sigma(T)$, $ \frac{m_s(T)}{\sigma(T)} $ , $ \frac{m_s(T)}{\sqrt{\sigma(T)}} $ and $ \frac{m_{0}(T)}{\sqrt{\sigma(T)}} $.\label{tabellamasse}}
\end{table}

\subsection{Concluding Remarks}

The main message of our analysis is that the mass of the lowest glueball (and thus likely the whole glueball spectrum) 
scales at finite temperature as $\sqrt{\sigma(T)}$ and thus is a $decreasing$ function of $T/T_c$ and vanishes at $T=T_c$.
Our results also suggest that the Isgur-Paton model is valid also at finite temperature and that its predicted scaling behaviour ($m_0\sim \sqrt{\sigma(T)}$)
can be conciliated with the different scaling behaviour predicted by the Svetitsky-Yaffe conjecture \cite{sy82} thanks to the appearance of a new mass scale
($m_s\sim \sigma(T)/T$) which in the vicinity of $T_c$ dominates the large $R$ behaviour of the correlator. This is likely to be a general mechanism. For instance,
a similar phenomenon was also observed a few years ago in the finite $T$ behaviour of the monopole spectrum of the random percolation gauge theory (see fig.2 of
ref.\cite{Giudice:2008uu}). It is also interesting to notice that this new mass scale strongly resembles the "spurious states" observed 
in \cite{Gregory:2012hu,Lucini:2010nv} which, in fact, were characterized by a large overlap with the torelon pair states 
 (we thank B. Lucini for this observation).

\noindent
It would be interesting to understand the physical meaning of $m_s$. Preliminary simulations show that at high enough temperatures the picture we have discussed 
holds almost unchanged even if we increase the distance between the two nearby Polyakov loops up to a few lattice spacings. In this limit our observable
describes the interaction of two mesons and, according to the effective string picture discussed above,  the mass scale $m_s$ should 
measure the attractive interaction between quarks and antiquarks belonging 
to different mesons.  Our results show that this
interaction becomes the dominant contribution in
the meson-meson correlator as $T_c$ is approached from below. This agrees with the intuitive
picture of deconfinement as a "melting" of mesons into individual quarks. 
In our framework this melting transition would be driven by the interaction mediated by the mass scale $m_s$.

{\bf Acnowledgements.} We thank F. Gliozzi, P. Grinza, B. Lucini, S. Lottini and P. Giudice for useful discussions and C. Caselle for help with the figures.

\end{document}

%% file: author_list.tex
\author{M.~Caselle} \affiliation{Dipartimento di Fisica Teorica dell'Universit\'a di Torino and I.N.F.N., via P.Giuria 1, I-10125 Torino, Italy }
 
\author{R.~Pellegrini} \affiliation{Dipartimento di Fisica Teorica dell'Universit\'a di Torino and I.N.F.N., via P.Giuria 1, I-10125 Torino, Italy }